\documentclass[12pt]{article}
\usepackage{amsfonts}
\usepackage{amssymb}
\usepackage{graphics,psboxit,amsmath}

\def\hybrid{\topmargin 10pt    \oddsidemargin 0pt
        \headheight 0pt \headsep 0pt
        \textwidth 6.35in       
        \textheight 9.25in       
        \marginparwidth .875in
        \parskip 5pt plus 1pt   \jot = 1.5ex}

\hybrid

\def\baselinestretch{1.2}

\catcode`\@=11

\def\marginnote#1{}
\newcount\hour
\newcount\minute
\newtoks\amorpm
\hour=\time\divide\hour by60
\minute=\time{\multiply\hour by60 \global\advance\minute by-\hour}
\edef\standardtime{{\ifnum\hour<12 \global\amorpm={am}%
        \else\global\amorpm={pm}\advance\hour by-12 \fi
        \ifnum\hour=0 \hour=12 \fi
        \number\hour:\ifnum\minute<10 0\fi\number\minute\the\amorpm}}
\edef\militarytime{\number\hour:\ifnum\minute<10 0\fi\number\minute}

\def\draftlabel#1{{\@bsphack\if@filesw {\let\thepage\relax
   \xdef\@gtempa{\write\@auxout{\string
      \newlabel{#1}{{\@currentlabel}{\thepage}}}}}\@gtempa
   \if@nobreak \ifvmode\nobreak\fi\fi\fi\@esphack}
        \gdef\@eqnlabel{#1}}
\def\@eqnlabel{}
\def\@vacuum{}
\def\draftmarginnote#1{\marginpar{\raggedright\scriptsize\tt#1}}

\def\draft{\oddsidemargin -.5truein
        \def\@oddfoot{\sl preliminary draft \hfil
        \rm\thepage\hfil\sl\today\quad\militarytime}
        \let\@evenfoot\@oddfoot \overfullrule 3pt
        \let\label=\draftlabel
        \let\marginnote=\draftmarginnote
   \def\@eqnnum{(\theequation)\rlap{\kern\marginparsep\tt\@eqnlabel}%
\global\let\@eqnlabel\@vacuum}  }

\def\preprint{\twocolumn\sloppy\flushbottom\parindent 2em
        \leftmargini 2em\leftmarginv .5em\leftmarginvi .5em
        \oddsidemargin -.5in    \evensidemargin -.5in
        \columnsep .4in \footheight 0pt
        \textwidth 10.in        \topmargin  -.4in
        \headheight 12pt \topskip .4in
        \textheight 6.9in \footskip 0pt
        \def\@oddhead{\thepage\hfil\addtocounter{page}{1}\thepage}
        \let\@evenhead\@oddhead \def\@oddfoot{} \def\@evenfoot{} }

\def\numberbysection{\@addtoreset{equation}{section}
        \def\theequation{\thesection.\arabic{equation}}}

\def\underline#1{\relax\ifmmode\@@underline#1\else
        $\@@underline{\hbox{#1}}$\relax\fi}

\def\titlepage{\@restonecolfalse\if@twocolumn\@restonecoltrue\onecolumn
     \else \newpage \fi \thispagestyle{empty}\c@page\z@
        \def\thefootnote{\fnsymbol{footnote}} }

\def\endtitlepage{\if@restonecol\twocolumn \else \newpage \fi
        \def\thefootnote{\arabic{footnote}}
        \setcounter{footnote}{0}}  

\catcode`@=12
\relax

\def\figcap{\section*{Figure Captions\markboth
        {FIGURECAPTIONS}{FIGURECAPTIONS}}\list
        {Figure \arabic{enumi}:\hfill}{\settowidth\labelwidth{Figure
999:}
        \leftmargin\labelwidth
        \advance\leftmargin\labelsep\usecounter{enumi}}}
 \relax
\def\tablecap{\section*{Table Captions\markboth
        {TABLECAPTIONS}{TABLECAPTIONS}}\list
        {Table \arabic{enumi}:\hfill}{\settowidth\labelwidth{Table
999:}
        \leftmargin\labelwidth
        \advance\leftmargin\labelsep\usecounter{enumi}}}
 \relax
\def\reflist{\section*{References\markboth
        {REFLIST}{REFLIST}}\list
        {[\arabic{enumi}]\hfill}{\settowidth\labelwidth{[999]}
        \leftmargin\labelwidth
        \advance\leftmargin\labelsep\usecounter{enumi}}}
 \relax
\makeatletter
\newcounter{pubctr}
\def\publist{\@ifnextchar[{\@publist}{\@@publist}}
\def\@publist[#1]{\list
        {[\arabic{pubctr}]\hfill}{\settowidth\labelwidth{[999]}
        \leftmargin\labelwidth
        \advance\leftmargin\labelsep
        \@nmbrlisttrue\def\@listctr{pubctr}
        \setcounter{pubctr}{#1}\addtocounter{pubctr}{-1}}}
\def\@@publist{\list
        {[\arabic{pubctr}]\hfill}{\settowidth\labelwidth{[999]}
        \leftmargin\labelwidth
        \advance\leftmargin\labelsep
        \@nmbrlisttrue\def\@listctr{pubctr}}}
 \relax
\makeatother
\newskip\humongous \humongous=0pt plus 1000pt minus 1000pt

\newif\ifdtup

\relax

\def\be{\begin{equation}}
\def\ee{\end{equation}}
\def\ba{\begin{eqnarray}}
\def\ea{\end{eqnarray}}

\def\a{\alpha}

\def\b{\beta}

\def\D{\Delta}

\def\no{\noindent}

\def\IR{\relax{\rm I\kern-.18em R}}
\def\II{\relax{\rm 1\kern-.35em1}}

\def\IR{\relax{\rm I\kern-.18em R}}
\def\inv{^{\raise.15ex\hbox{${\scriptscriptstyle -}$}\kern-.05em 1}}

\begin{document}

\begin{titlepage}
\begin{center}

\hfill IFT-UAM/CSIC-13-006\\
\vskip -.1 cm

\vskip 0.4in

{\LARGE Holographic correlation functions of hexagon Wilson loops with one local operator}
\vskip 0.4in

{\bf Rafael Hern\'andez$^1$}\phantom{x} and\phantom{x}
{\bf Juan Miguel Nieto}$^2$ 
\vskip 0.1in

${}^1\!$
Departamento de F\'{\i}sica Te\'orica I\\
Universidad Complutense de Madrid\\
$28040$ Madrid, Spain\\
{\footnotesize{\tt rafael.hernandez@fis.ucm.es}}

\vskip .2in

${}^2\!$
Instituto de F\'{\i}sica Te\'orica UAM/CSIC \\
Universidad Aut\'onoma de Madrid \\
$28049$ Madrid, Spain\\
{\footnotesize{\tt juanmiguel.nieto@estudiante.uam.es}}

\end{center}

\vskip .2in

\centerline{\bf Abstract}
\vskip .1in
\no

\noindent
We consider the ratio of the correlation function of an hexagon light-like Wilson loop with one local operator over the expectation 
value of the Wilson loop within the strong-coupling regime of the AdS/CFT correspondence. 
We choose the hexagon Wilson loop within a class of minimal solutions obtained by cutting and gluing light-like quadrangle loops. 
These surfaces do not have an interpretation in terms of dual scattering amplitudes but they still exhibit general features 
of the mixed correlation function. 
In the case of a regular null hexagon conformal symmetry constrains the space-time dependence of the correlator up to a function 
of three conformal cross-ratios. We obtain the leading-order contribution to the correlation function in the semiclassical approximation 
of large string tension, and express the result in terms of three conformal ratios in the case where the local operator is taken to be 
the dilaton. We include the analysis of an irregular Wilson loop obtained after a boost of the regular hexagon.

\vskip .4in
\noindent

\end{titlepage}
\vfill
\eject

\def\baselinestretch{1.2}


\baselineskip 20pt


\no
The study of Wilson loop observables within the AdS/CFT correspondence has lead to impressive progress on our understanding 
of the duality.  One of the most insightful developments came from the observation that the expectation 
values of light-like polygonal Wilson loops in planar Yang-Mills with ${\cal N}=4$ supersymmetry can be related to gluon scattering amplitudes~\cite{AM}. 
The proposal, originally suggested in the strong-coupling regime, was also shown to hold at weak-coupling~\cite{weakcoupling}.
More recently light-like polygonal Wilson loops were also identified with correlation functions of local gauge invariant primary 
operators~\cite{correlationWilson}. In the limit where the insertion points of the local operators become pairwise null-separated 
the correlation function becomes proportional to the expectation value of a light-like Wilson loop in the adjoint 
representation, with cusps at the position of the primary operators. 
This conjecture was extended in \cite{ABT} (see also references \cite{ER}, \cite{Adamo}) to mixed correlation 
functions of null Wilson loops with some local operators. In particular, the correlation function of $n+1$ operators located at pairwise 
null-separated points should be related to the correlation function of an $n$-sided light-like Wilson loop with one local operator,
\be
\hbox{lim}_{\, |x^{(i,i+1)}| \ \rightarrow \ 0} 
\frac {\langle {\cal O} (x^{(1)}) \ldots {\cal O} (x^{(n)}) {\cal O} (a) \rangle}{\langle {\cal O} (x^{(1)}) \ldots {\cal O} (x^{(n)}) \rangle} 
\sim \frac {\langle W_n {\cal O} (a) \rangle}{\langle W_n \rangle} = {\cal C}(W_n,a) \ .
\ee

\no
The analysis of mixed correlation functions is of great importance because they can be employed to extract information 
about the operator product expansion of the Wilson loop. An extensively studied example is that of correlation functions of circular 
Wilson loops with various types of local operators \cite{Berenstein}-\cite{Enari}. In these cases conformal symmetry constrains 
the dependence of the correlation function on the space-time position 
of the local operators. On the contrary, symmetries do not suffice to fix completely the correlation functions of null Wilson loops 
and primary operators. In \cite{ABT} it was argued how conformal symmetry constrains 
the ratio ${\cal C}(W_n,a)$ to depend on the $n$ distances $|a-x^{(i)}|$ from the scalar operator to the positions of the cusps at $x^{(i)}$, 
the $n(n-3)/2$ non-vanishing diagonals of the polygon $|x^{(i)}-x^{(j)}|$, where $i \neq j \pm 1$, and an undetermined function $F$ 
of $3n-11$ conformal cross-ratios and the 't Hooft coupling, 
\be
{\cal C}(W_n,a) = \frac {\prod_{i<j-1}^{n}|x^{(i)}-x^{(j)}|^{\frac {2\Delta}{n(n-3)}}}{\prod_{i=1}^n |a-x^{(i)}|^{\frac {2 \Delta}{n}}} 
F(\zeta_1, \ldots , \zeta_{3n-11}) \ .
\label{structureCn}
\ee
The structure of the conformal function $F$ was analyzed in reference \cite{ABT} in the first non-trivial case which is that 
of the correlation function of a quadrangle Wilson loop and one local operator, depending on just a single cross-ratio, 
\be
\zeta = \frac {|a-x^{(2)}|^2 |a-x^{(4)}|^2 |x^{(1)}-x^{(3)}|^2}{|a-x^{(1)}|^2 |a-x^{(3)}|^2 |x^{(2)}-x^{(4)}|^2} \ .
\ee
The leading order contribution to $F(\zeta)$ was obtained both in the strong and the weak-coupling regimes 
with the local operator taken as the dilaton or the chiral primary scalar~\cite{ABT}. 
In the case of a lagrangian operator the next-to-leading order term in the weak-coupling expansion 
was found in~\cite{AHS}, and the two-loop contribution has been recently computed in~\cite{AHS2}. 
The analysis of higher orders both in the weak and the strong-coupling expansions is a very interesting 
question and remains a challenging problem. 

\no
Another interesting direction to explore is the extension to the case of polygons with a larger number of cusps. Symmetry 
considerations were employed in \cite{ABT} to prove the general structure in expression (\ref{structureCn}) for the correlation 
function of a light-like pentagon with a local operator, although a strong-coupling analysis could not be included in this case because 
the world-sheet surface ending on a Wilson loop with five cusps is not known. The next case of interest is that of a null hexagon, 
where the dependence of the correlation function on the position of the cusps and the location of the scalar operator should be 
determined up to a function of $3 \times 6 - 11 = 7$ independent conformal ratios $\zeta_k$, 
\be
{\cal C}(W_{6},a) = \frac {\prod_{i<j-1}^{6}|x^{(i)}-x^{(j)}|^{\frac {\Delta}{9}}}{\prod_{i=1}^6 |a-x^{(i)}|^{\frac {\Delta}{3}}} 
F(\zeta_1, \zeta_2 , \ldots , \zeta_7) \ .
\label{C6prediction}
\ee

\no
In general finding the function $F$ for a general null hexagon with nine different diagonals seems a complicated problem. 
But an important simplification arises when the hexagon is regular. For this class of Wilson loops only three cross-ratios 
are independent and the structure of the conformal function $F$ simplifies greatly. \footnote{As we will prove below, no additional structure arises 
when the hexagon is deformed by performing a boost transformation and the corresponding irregular correlation function can 
still be written as a function of three conformal cross-ratios.} At weak-coupling this function was obtained perturbatively  
for a regular null hexagon with one dilaton operator in \cite{ABT}. However an analysis of the correlation function at strong-coupling was again 
not possible because there is no explicit solution for a minimal surface 
ending on a light-like hexagon dual to the scattering amplitude of six gluons. \footnote{Integrability of the underlying string sigma 
model was employed in \cite{int} to compute scattering amplitudes at strong-coupling without explicit knowledge of the dual minimal surface.} 
In this note we will consider instead some other minimal solutions ending in a null hexagon. In particular we will analyze the minimal surfaces constructed 
in \cite{hexagon} by cutting and gluing the original quadrangle solutions in reference~\cite{AM}. 
The class of solutions obtained by this procedure does not have a clear interpretation as scattering amplitudes because in order 
to derive them some additional boundary conditions need to be imposed, and as a consequence the resulting surfaces contain only a part of the 
diagrams that contribute to the complete six-point amplitudes. But they correspond still to minimal surfaces ending on a light-like 
hexagonal Wilson loop and thus they can be employed to analyze the semiclassical contribution to the strong-coupling limit 
of the correlation function of a null hexagon with one local operator. 

Before proceeding we will first briefly discuss how the correlation function 
of a polygonal Wilson loop with one local operator can be evaluated within the semiclassical approximation. 
Let us first recall that at strong-coupling the local operator can be described by means of a marginal vertex operator integrated 
over the string world-sheet,
\be
{\cal O}(a) = \int d^2 \xi \, V[X(\xi), a] \ .
\label{vertex}
\ee
Therefore we will be interested in the evaluation of the correlation function
\be
\langle W_n {\cal O}(a) \rangle = \int DX \, e^{-I[X]} \int d^2 \xi \, V[X(\xi), a] \ ,
\ee
where $DX$ is the integration measure over the embedding coordinates of the string world-sheet and $I[X]$ is the string action. 
In the planar limit the integration is performed over disc-like euclidean world-sheets ending on the null Wilson loop. 
In order to evaluate the integral we will assume that the local operator 
carries finite charges. In the semiclassical limit of large string tension $\sqrt{\lambda} \gg 1$ the contribution from the {\em light} vertex operator 
to the stationary surface in the string path integral can be ignored and 
it is the Wilson loop that dominates the correlation function. The leading contribution to the correlation function is then given by
\be
\langle W_n {\cal O}(a) \rangle = \langle W_n \rangle \int d^2 \xi \, \hat{V}[X(\xi), a] \ ,
\ee
where $\hat{V}[X,a]$ denotes the vertex operator evaluated on the Wilson loop surface. 
Therefore at strong-coupling the ratio ${\cal C}(W_n,a)$ can be obtained upon integration 
of the light vertex operator over the classical world-sheet solution. 

We will now discuss some relevant aspects of the minimal world-sheet surfaces ending in a null hexagon that we will employ 
in this note. In Poincar\'e coordinates $z$ and $x=(x_0,x_1,x_2,x_3)$ the surfaces 
that we will consider will be solutions satisfying the condition~\footnote{The analysis 
that we will present below can be extended to other minimal surfaces constructed in \cite{hexagon}, 
such as those satisfying the condition 
\(
x_0 (x_1,x_2) = \frac {1}{2} \big( |x_1 x_2| + x_1 x_2 - |x_1| x_2 + x_1 |x_2| \big) .
\)
However, the Wilson loop at the boundary in this case is slightly less symmetrical because the barycenter of the hexagon does not coincide 
with the coordinate origin. As a consequence the corresponding correlation functions become much more involved.}
\be
x_0 (x_1,x_2) = -x_1 |x_2| \ .
\ee 
The euclidean world-sheet in $AdS_5$ ending in the light-like hexagon at the boundary is \cite{hexagon}
\ba
z & \!\! = \!\! & \frac {\a}{\cosh u \cosh v \pm \b \sinh u \sinh v} \ , \quad 
x_0 = \frac {\pm \a \sqrt{1+\b^2} \sinh u \sinh v}{\cosh u \cosh v \pm \b \sinh u \sinh v} \ , \nonumber \\
x_1 & \!\! = \!\! & \frac{\a \sinh u \cosh v}{\cosh u \cosh v \pm \b \sinh u \sinh v} \ , \quad 
x_2 = \frac{\a \cosh u \sinh v}{\cosh u \cosh v \pm \b \sinh u \sinh v} \ , \label{Poincarehexagon}
\ea
while $x_3$ can be taken to vanish. The variables $u,v  \in (-\infty,+\infty)$ parameterize the minimal surface, 
the constant $\alpha$ measures the extension of the Wilson loop and $\beta$ is a boost parameter deforming the 
shape of the hexagon. The minus signs in expressions (\ref{Poincarehexagon}) are meant when $v>0$ 
while the plus signs correspond to the region where $v<0$. The vertices of the null hexagon at the boundary are located at 
\begin{gather}
x^{(1)} = \frac {1}{1 - \beta} \big( - \a \sqrt{1+\b^2}, \a,  \a, 0 \big) \ , \quad
x^{(2)} = \frac {1}{1 + \b} \big( \, \a \sqrt{1+\b^2}, - \, \a, \a, 0 \big) \ , \nonumber \\
x^{(3)} = \big( 0, -\, \a, 0, 0 \big) \ , \quad 
x^{(4)} = \frac {1}{1 + \b} \big( \, \a \sqrt{1+\b^2}, -\, \a, -\, \a, 0 \big) \ , \label{hexagon} \\
x^{(5)} = \frac {1}{1 - \b} \big( - \a \sqrt{1+\b^2}, \, \a, - \a, 0 \big) \ , \quad 
x^{(6)} = \big( 0, \a, 0, 0 \big) \ . \nonumber
\end{gather}
It is immediate to check that they define indeed an irregular null polygon with six light-like sides 
and six different non-vanishing diagonals,
\ba
|x^{(1)}-x^{(3)}|^2 & \!\!\! = \!\!\! & |x^{(3)}-x^{(5)}|^2 = \frac {4\a^2}{1-\b} \ , \quad \quad \:
|x^{(1)}-x^{(5)}|^2 = \frac {4\a^2}{(1-\b)^2} \ , \nonumber \\
|x^{(1)}-x^{(4)}|^2 & \!\!\! = \!\!\! & |x^{(2)}-x^{(5)}|^2 = \frac {4\a^2}{1-\b^2} \ , \quad \quad 
|x^{(2)}-x^{(4)}|^2 = \frac {4\a^2}{(1+\b)^2} \ , \label{diagonales} \\
|x^{(2)}-x^{(6)}|^2 & \!\!\! = \!\!\! & |x^{(4)}-x^{(6)}|^2 = \frac {4\a^2}{1+\b} \ , \quad \quad \:\:
|x^{(3)}-x^{(6)}|^2 = 4 \a^2 \ . \nonumber 
\ea
Setting the boost to zero corresponds to a regular hexagon with equal length diagonals. 

We will now proceed to evaluate the correlation function of this hexagon Wilson loop with a light dilaton operator. 
Other choices of light vertex such as the superconformal chiral primary operator are of course possible and deserve a separate 
study. At large string tension the leading contribution to the massless dilaton vertex is purely bosonic,
\be
V^{\hbox{\tiny{(dilaton)}}} (a) = 
c_{\D} \, \int d^2 \xi \, K_{\D}(z;x,a) \, \big( \! \cos \theta e^{i \varphi} \big)^j\, 
\big[ z^{-2} (\partial x_m \bar{\partial}x^m + \partial z \bar{\partial} z ) + \partial X_K \bar{\partial} X_K \big] \ ,
\label{Vdilaton}
\ee
where $c_{\D}$ is the normalization constant of the dilaton \cite{Berenstein,ZaremboRT},
\be
c_{\D} = \frac {\sqrt{\lambda}}{8 \pi N} \sqrt{(j+1)(j+2)(j+3)} \ ,
\ee 
$K_{\D}(z;x,a)$ is the bulk-to-boundary propagator,  
\be
K_{\D}(z;x,a) =  \left[ \frac {z}{z^2+(x-a)^2} \right]^{\Delta} \ ,
\ee
and the derivatives are $\partial \equiv \partial_+$ and $\bar{\partial} \equiv \partial_-$. To leading order the scaling dimension 
at strong-coupling is $\Delta=4+j$, where $j$ is the Kaluza-Klein momentum of the dilaton. 
The corresponding dual gauge invariant operator is $\hbox{Tr}(F_{\mu \nu}^2 Z^j + \cdots )$. 
For simplicity in this note we will focus on the $j=0$ case, so that the coupling is just to the lagrangian operator. 

\subsection*{Regular hexagon}

We will first analyze the case of the regular hexagon. 
As discussed above in order to find the semiclassical contribution to the correlation function 
we need to evaluate the dilaton vertex operator~(\ref{Vdilaton}) in the solution (\ref{Poincarehexagon}) for the null hexagon with $\beta=0$. We find
\be
{\cal C} (W_6,a) = 2 c_{\D}  \int du dv  \left[ \frac {(\cosh u \, \cosh v)^{-1}}{q-2a_1 \tanh u-2a_2 \tanh v\pm 2a_0 \tanh u \tanh v} \right]^{4} \ ,
\label{intC6}
\ee
where $(a_0,a_1,a_2,a_3)$ are the coordinates of the point $a$, and we have defined 
\be
q=1-a_0^2+a_1^2+a_2^2+a_3^2 \ . 
\label{q}
\ee
The minus sign in the integrand in (\ref{intC6}) corresponds again to $v>0$ while the plus refers to the region where $v<0$. 
Note that to alleviate expressions below we have also set the size $\a$ to one. Evaluating the integral is straightforward 
if following reference \cite{ABT} we change variables to $U = \tanh u$ and $V= \tanh v$, and split the integration interval to take 
care of the signs in~(\ref{intC6}). We find
\begin{align}
&{\cal C}(W_6,a) = \frac {c_{\D}}{3} \left[ 
\cfrac{8\big( q^2 \, (a_0^2 a_1^2 - a_1^2 a_2^2 - a_0^3 a_1 + 2 a_0 a_1 a_2^2) + 4a_1^4 a_2^2 - 4 a_0 a_1^3 a_2^2 \big) - 2 q^4 a_0^2}
{(q^2 - 4a_1^2) (q^2 a_0^2-4a_1^2 a_2^2)^2} \right. \label{CW6} \\
&\left. + \, \cfrac{q^2+4(a_0^2-a_1^2-a_2^2)}{4 (q^2 a_0^2 -4 a_1^2 a_2^2)^3} 
\Big[ \, (12 a_0 a_1^2 a_2^2 q+a_0^3 q^3 ) \log (q_1^2 q_2 q_3) + ( 8 a_1^3 a_2^3 + 6 a_0^2 a_1 a_2 q^2 ) 
\log \Big( \frac {q_2}{q_3} \Big) \Big] \right] \nonumber \ ,
\end{align}
where we have introduced the quotients
\be
q_1 = \frac{q-2a_1}{q+2a_1} \ , \quad q_2 = \frac {q+2(a_0+a_1+a_2)}{q-2(a_0+a_1-a_2)} \ , \quad
q_3 = \frac {q+2(a_0+a_1-a_2)}{q-2(a_0+a_1+a_2)} \ .
\label{q1q2q3}
\ee
This result is in agreement with the expected general behavior (\ref{C6prediction}) for the correlation function 
of a light-like regular hexagon and one local operator. In order to exhibit this let us first present a basis of independent 
conformal cross-ratios. 

\no
The cross-ratios are conformally-invariant combinations of the six vertices of the hexagon and the point $a$ 
where the dilaton operator is located. Therefore they involve both the diagonals of the null hexagon 
and the six distances $|a-x^{(i)}|$. 
In general, seven conformal ratios can be constructed from the nine different diagonals of an irregular hexagon. But when 
some of the diagonals have equal length some cross-ratios can be written in terms of others. 
In the case of a regular null hexagon only three conformal cross-ratios are independent. A possible choice is that in~\cite{ABT},
\begin{gather}
\zeta_1 = \frac {|a-x^{(5)}|^2 |x^{(1)}-x^{(3)}|^2}{ |a-x^{(3)}|^2 |x^{(1)}-x^{(5)}|^2} \ , \quad 
\zeta_2 = \frac {|a-x^{(6)}|^2 |x^{(2)}-x^{(4)}|^2}{ |a-x^{(4)}|^2 |x^{(2)}-x^{(6)}|^2} \ ,  \nonumber \\
\zeta_3 = \frac {|a-x^{(3)}|^2 |a-x^{(6)}|^2 |x^{(1)}-x^{(4)}|^2}{ |a-x^{(1)}|^2 |a-x^{(4)}|^2 |x^{(3)}-x^{(6)}|^2} \ . \quad  \label{threeconformalratios} 
\end{gather}
If we write now the distances from the dilaton operator to the six vertices in terms of $q$ 
we immediately recognize the various factors in expressions (\ref{q1q2q3}), 
\begin{gather}
|a-x^{(1)}|^2 = q - 2 (a_0 + a_1 + a_2) \ , \quad |a-x^{(2)}|^2 = q + 2 (a_0 + a_1 - a_2) \ , \nonumber \\
|a-x^{(3)}|^2 = q + 2 a_1 \ , \quad |a-x^{(4)}|^2 = q + 2 (a_0 + a_1 + a_2) \ , \label{axdistances} \\
|a-x^{(5)}|^2 = q - 2 (a_0 + a_1 - a_2) \ , \quad |a-x^{(6)}|^2 = q - 2 a_1 \ . \nonumber
\end{gather}
Noting that these distances satisfy $|a-x^{(1)}|^2 + |a-x^{(4)}|^2 = 2q$, and cyclic, 
it is a simple exercise to express the quotients (\ref{q1q2q3}) in terms of the three conformal ratios (\ref{threeconformalratios}),
\be
q_1 = \frac {\zeta_2 (\zeta_2 - \zeta_3)}{\zeta_3 (\zeta_2 -1)} \ , \quad q_2 =\frac{\zeta_2 - \zeta_3}{\zeta_1 \zeta_3 (\zeta_2 -1)} \ , \quad
q_3 = \frac {\zeta_2^2 + \zeta_3 (\zeta_1 - 1) - \zeta_1 \zeta_2 \zeta_3}{\zeta_2 (\zeta_2 - 1)} \ .
\ee
In a similar manner after some algebra all factors in equation (\ref{CW6}) can be written as 
\begin{align}
&a_0 =\frac {qP_1}{2 P_0} \ , \quad a_1 a_2 = \frac {q^2 P_2}{4 P_0^2} \ ,\quad a_2 =\frac {q P_3}{2 P_0} \ ,\nonumber \\
&q^2-4 a_1^2 =\frac {4q^2 P_4}{P_0^2} \ , \quad
q^2+4(a_0^2-a_1^2-a_2^2) =  \frac {2q^2 P_5}{P_0^2} \ , 
\label{Pdefinition}
\end{align}
where the $P_i$'s are some polynomial functions of the conformal cross-ratios given by
\ba
P_0 & \!\!\! =  \!\!\! & \zeta_2^2 - \zeta_3 \ , \quad 
P_1 = \zeta_1 \zeta_3 +\zeta_2 \big( \zeta_2-\zeta_3 (\zeta_1 + 2) + 1 \big) \ , \nonumber \\
P_2 & \!\!\! =  \!\!\! & (\zeta_2 - \zeta_1 \zeta_3) (\zeta_2 - 1) (\zeta_2^2 + \zeta_3 - 2 \zeta_2 \zeta_3) \ , \quad
P_3 = (\zeta_1 \zeta_3 - \zeta_2) (\zeta_2 - 1) \ , \\
P_4 & \!\!\! =  \!\!\! & \zeta_2 \zeta_3 (\zeta_2 - \zeta_3) (\zeta_2 - 1) \ , \quad 
P_5 = 2\zeta_2 \big[ \zeta_2^2 + \zeta_2 \zeta_3\big( \zeta_2 (\zeta_3-1) - 2 \big) + \zeta_3 \big( \zeta_1 -\zeta_1 \zeta_3 +\zeta_3 \big) \big] \ . \nonumber
\ea
It suffices to insert expressions (\ref{Pdefinition}) in equation (\ref{CW6}) to prove that the correlation function of a regular 
hexagon Wilson loop with light-like sides and one lagrangian operator can be written as a function of three conformal cross-ratios. 
Equation (\ref{CW6}) is indeed of the general form (\ref{C6prediction}), where the conformal function $F=F(\zeta_1 ,\zeta_2 ,\zeta_3)$ is given by
\begin{align}
&F(\zeta_1 ,\zeta_2 ,\zeta_3) = 
\frac {c_{\D}}{24} \frac{\big(\zeta_1^3 \zeta_2^3 \zeta_3^4 q_2^2 q_3 \big)^{4/6} \big( \zeta_2 -1 \big)^{4}}{P_4 ( P_0^2 P_1^2 - P_2^2 )^2}  \nonumber \\
& \times \Big\{P_0^2 \big( P_2^2 + 2 P_1 P_2 P_3 + P_1^2 ( P_2 +4 P_0^2  - 20 P_4 ) \big) 
+ P_2^2 \big(P_2 - 8 P_4 \big) \label{Fhexagon} \\ 
& + \frac{2 P_4 P_5}{( P_0^2 P_1^2 - P_2^2 )} \Big[ P_0 P_1 (3 P_2^2 + P_0^2 P_1^2) \log (q_1^2 q_2 q_3) 
+ P_2 (3 P_0^2 P_1^2 + P_2^2) \log \Big( \frac {q_2}{q_3} \Big) \Big] \Big\} \ . \nonumber
\end{align}

Now there are various limits of interest that we may consider. A possible situation is the limit where the lagrangian operator 
is located near a cusp, that is, when $a=x^{(i)}+\epsilon \alpha$ with $\epsilon \rightarrow 0$ and $|\alpha|=1$. 
For instance when $a \rightarrow x^{(1)}$ we find
\ba
{\cal C}(W_6,a) _{a \rightarrow x^{(1)}}  
& \!\!\! = \!\!\! & -\frac{ c_{\D}}{3 \epsilon^4} \left[ \frac{5\alpha_0 +\alpha_1+4\alpha_2}{2 (\alpha_1 + \alpha_2 ) 
(1-2\alpha_0^2 -2\alpha_0 \alpha_1 -2\alpha_0 \alpha_2 -2\alpha_1 \alpha_2 )^2} \right. \\
& \!\!\! + \!\!\! &\left.  \frac{1 +2\alpha_0^2 +2\alpha_0 \alpha_1 +2\alpha_0 \alpha_2 +2\alpha_1 \alpha_2 }
{ (1-2\alpha_0^2 -2\alpha_0 \alpha_1 -2\alpha_0 \alpha_2 -2\alpha_1 \alpha_2)^3 } 
\log \left[ 2(\alpha_0+\alpha_1)(\alpha_0+\alpha_2) \right] \right] \ . \nonumber
\ea
This is indeed the expected behavior from the general structure of the correlator,~\footnote{Some care 
needs to be taken in order to recover this dependence from expression (\ref{C6prediction}), because the conformal ratios 
scale differently as the lagrangian approaches one of the vertices of the null hexagon. In the case at hand $\zeta_1$ remains finite, 
but $\zeta_2 \sim {\cal O}(\epsilon)$ while $\zeta_3 \sim {\cal O}(1/\epsilon)$. Inserting this dependence in (\ref{C6prediction}) 
and using the conformal function (\ref{Fhexagon}) we recover indeed the general behavior below.} 
\be
{\cal C}(W_6,a) _{a \rightarrow x^{(1)}}  \sim \frac{1}{|a-x^{(1)}|^4} \ .
\ee
Another limit of interest is that where instead of approaching one of the cusps the lagrangian operator is null-separated from it. 
In this case the correlation function develops a logarithmic singularity. If we take for instance $|a - x^{(1)}| \rightarrow 0$ we find
\be
{\cal C}(W_6,a) _{|a - x^{(1)}| \, \rightarrow \, 0} \sim \log |a - x^{(1)}| \ .
\ee
When the lagrangian is null-separated simultaneously from two adjacent vertices we also find a singular behavior. 
Now the divergence comes from the factors $1/(q^2a_0^2 -4 a_1^2 a_2^2)^2$ and $(q^2+4(a_0^2-a_1^2-a_2^2))/4(q^2 a_0^2 -4a_1^2 a_2^2)^3$ 
in expression (\ref{CW6}). Taking for instance the limit $|a - x^{(1)}| \, , |a - x^{(2)}| \rightarrow 0$ we find
\be
{\cal C}(W_6,a) _{|a - x^{(1)}| \, , \, |a-x^{(2)}| \, \rightarrow \, 0} \sim \frac {1}{|a-x^{(1)}|^2 |a-x^{(2)}|^2} \ .
\ee
This seems to be a general scaling, exhibited also by the quadrangle Wilson loop in \cite{ABT}.
We could also consider the case where the operator is located 
far away from the null hexagon. This limit provides the coefficient of the lagrangian in the operator product expansion of the Wilson loop. 
However in this limit there is no way to distinguish the hexagon (\ref{Poincarehexagon}) from the quadrangle 
solution. Therefore taking $|a| \rightarrow \infty$ 
we find
\be
{\cal C}(W_6,a) _{|a|\rightarrow \infty}=\frac{32 c_{\D}}{9|a|^8} \ ,
\ee
rather than the predicted coefficient upon numerics in \cite{ABT}. This result is an artifact of the cutting and gluing derivation of the 
hexagon Wilson loop under consideration.

The analysis in the previous paragraphs can also be extended to general values of the scaling dimension. Finding the correlation 
function of a null hexagon with a general dilaton requires evaluating an integral as in equation (\ref{intC6}), with the exponent in the 
propagator replaced by $\D=4+j$. This integral can be written in terms of analytic functions if we restrict the computation to some convenient 
locations of the dilaton operator. However the resulting expressions add little to our discussion and we will not present them here.


\subsection*{Irregular hexagon}

Now we will extend the calculation to the case of an irregular hexagon with non-vanishing~$\b$. 
When we evaluate the lagrangian vertex operator in the general hexagon we get
\be
{\cal C}(W_6,a) = 2
c_{\D}  \int du dv  \left[ \frac {(\cosh u \, \cosh v)^{-1}}{q-2a_1 \tanh u-2a_2 \tanh v\pm 2\tilde{a}_0 \tanh u \tanh v} \right]^{4} \ ,
\label{intC6irregular}
\ee
with $q$ as defined in (\ref{q}) and the integration interval as in (\ref{intC6}). 
This integral is identical to the one for the regular hexagon, except for the deformed component $\tilde{a}_0$,  
\be
\tilde{a}_0=a_0 \sqrt{1-\beta^2} + \frac{q-2}{2} \beta \ ,
\ee
that collects the effect of the boost on the null hexagon. The expression for the correlation function is therefore the same as in the case of the regular 
Wilson loop, but with the coordinate $a_0$ replaced by $\tilde{a}_0$ in equation~(\ref{CW6}) and the quotients (\ref{q1q2q3}). As before, in order to write 
the correlation function in terms of conformal cross-ratios we will first write the distances $|a-x^{(i)}|$ in terms of q. Now we find
\begin{gather}
|a-x^{(1)}|^2 = \frac {\alpha}{1-\beta} \big( q - 2 (\tilde{a}_0 + a_1 + a_2) \big) \ , \quad 
|a-x^{(2)}|^2 = \frac {\alpha}{1+\beta} \big(q + 2 (\tilde{a}_0 + a_1 - a_2) \big) \ , \nonumber \\
|a-x^{(3)}|^2 = \alpha \big(q + 2 a_1 \big) \ , \quad 
|a-x^{(4)}|^2 = \frac {\alpha}{1+\beta} \big( q + 2 (\tilde{a}_0 + a_1 + a_2) \big) \ , \\
|a-x^{(5)}|^2 = \frac {\alpha}{1-\beta} \big( q - 2 (\tilde{a}_0 + a_1 - a_2) \big) \ , \quad 
|a-x^{(6)}|^2 = \alpha \big( q - 2 a_1 \big) \ . \nonumber
\end{gather}
When we combine these expressions with the length of the diagonals~(\ref{diagonales}) we realize that
\begin{equation}
\left. \frac{|a-x^{(i)}| |a-x^{(j)}|}{|x^{(i)}-x^{(j)}|} \right|_{\beta\neq 0}=\left. \frac{|a-x^{(i)}| |a-x^{(j)}|}{|x^{(i)}-x^{(j)}|} \right|_{\beta = 0 \, , \, a_0\rightarrow \tilde{a}_0} \ .
\end{equation}
Therefore once we write the conformal ratios in terms of the coordinates of $a$ they appear to have 
the same functional dependence as in the regular case, with the lagrangian located at $\tilde{a}=(\tilde{a}_0,a_1,a_2,a_3)$. 
Thus the boost parameter drops from the three cross-ratios and the function $F(\zeta_1 ,\zeta_2 ,\zeta_3)$ 
agrees in the case of the irregular hexagon with expression (\ref{Fhexagon}) for the regular Wilson loop, 
with the dilaton shifted to the point $\tilde{a}$. The absence of additional structure in the deformed Wilson loop 
is a consequence of the fact that it has been constructed by applying a conformal transformation to the regular hexagon.

There are many possible extensions of the problem addressed in this note. A natural one 
is the analysis of the general class of null polygons dual to gluon scattering amplitudes. But as already stated above no explicit 
solutions are known beyond the case of the Wilson loop with four cusps. However in spite of the absence of explicit solutions 
a general description of regular light-like polygons with an even number of sides based on Pohlmeyer reduction of the equations 
of the string was introduced in \cite{AMeven}. This description inspired later one a general method to obtain the areas of the corresponding 
minimal surfaces with no need to know their shape~\cite{int}. It would be very interesting if that approach could be extended 
to evaluate mixed correlation functions at strong-coupling. 


\vspace{8mm}

\centerline{\bf Acknowledgments}

\vspace{2mm}

\no
R.~H. wishes to thank the Theory Division at CERN for hospitality and support at the early stages of this work. 
The work of R.~H. is supported by MICINN through a Ram\'on y Cajal contract and grant FPA2011-24568, 
and by BSCH-UCM through grant GR58/08-910770. 



\begin{thebibliography}{99}

\renewcommand{\baselinestretch}{.99}
\normalsize

\bibitem{AM} L.~F.~Alday and J.~M.~Maldacena,
{\em Gluon scattering amplitudes at strong coupling},
JHEP {\bf 0706} (2007) 064, {\tt [arXiv:0705.0303 [hep-th]]}.

\bibitem{weakcoupling} G.~P.~Korchemsky, J.~M.~Drummond and E.~Sokatchev,
{\em Conformal properties of four-gluon planar amplitudes and Wilson loops}, 
Nucl.\ Phys.\ B {\bf 795} (2008) 385, {\tt [arXiv:0707.0243 [hep-th]]}.

A.~Brandhuber, P.~Heslop and G.~Travaglini,
{\em MHV amplitudes in ${\cal N}=4$ super Yang-Mills and Wilson loops},
Nucl.\ Phys.\ B {\bf 794} (2008) 231, {\tt [arXiv:0707.1153 [hep-th]]}.

\bibitem{correlationWilson} L.~F.~Alday, B.~Eden, G.~P.~Korchemsky, J.~Maldacena and E.~Sokatchev,
{\em From correlation functions to Wilson loops},
JHEP {\bf 1109} (2011) 123, {\tt [arXiv:1007.3243 [hep-th]]}.

\bibitem{ABT} L.~F.~Alday, E.~I.~Buchbinder and A.~A.~Tseytlin,
{\em Correlation function of null polygonal Wilson loops with local operators},
JHEP {\bf 1109} (2011) 034, {\tt [arXiv:1107.5702 [hep-th]]}.

\bibitem{ER} O.~T.~Engelund and R.~Roiban,
{\em On correlation functions of Wilson loops, local and non-local operators},
JHEP {\bf 1205} (2012) 158, {\tt [arXiv:1110.0758 [hep-th]]}.
  
\bibitem{Adamo} T.~Adamo,
{\em Correlation functions, null polygonal Wilson loops, and local operators},
JHEP {\bf 1112} (2011) 006, {\tt [arXiv:1110.3925 [hep-th]]}.

\bibitem{Berenstein} D.~E.~Berenstein, R.~Corrado, W.~Fischler and J.~M.~Maldacena,
{\em The Operator product expansion for Wilson loops and surfaces in the large N limit},
Phys.\ Rev.\ D {\bf 59} (1999) 105023, {\tt [arXiv:hep-th/9809188]}.

\bibitem{DGO} N.~Drukker, D.~J.~Gross and H.~Ooguri,
 {\em Wilson loops and minimal surfaces},
Phys.\ Rev.\  D {\bf 60} (1999) 125006, {\tt [arXiv:hep-th/9904191]}.

\bibitem{ESZ} J.~K.~Erickson, G.~W.~Semenoff and K.~Zarembo,
{\em Wilson loops in ${\cal N}=4$ supersymmetric Yang-Mills theory},
Nucl.\ Phys.\ B {\bf 582} (2000) 155, 
{\tt [arXiv:hep-th/0003055]}.

N.~Drukker and D.~J.~Gross,
{\em An Exact prediction of ${\cal N}=4$ SUSYM theory for string theory},
J.\ Math.\ Phys.\  {\bf 42} (2001) 2896, 
{\tt [arXiv:hep-th/0010274]}.

\bibitem{ZaremboW} K.~Zarembo,
{\em Open string fluctuations in $AdS_5 \times S^5$ and operators with large R charge},
Phys.\ Rev.\  D {\bf 66} (2002) 105021, {\tt [arXiv:hep-th/0209095]}.
  
\bibitem{Pestun} V.~Pestun and K.~Zarembo,
{\em Comparing strings in $AdS_5 \times S^5$ to planar diagrams: An example},
Phys.\ Rev.\  D {\bf 67} (2003) 086007, {\tt [arXiv:hep-th/0212296]}.

\bibitem{Tsuji} A.~Tsuji,
{\em Holography of Wilson loop correlator and spinning strings},
Prog.\ Theor.\ Phys.\  {\bf 117} (2007) 557, {\tt [arXiv:hep-th/0606030]}.

A.~Miwa and T.~Yoneya,
{\em Holography of Wilson-loop expectation values with local operator insertions},
JHEP {\bf 0612} (2006) 060, {\tt [arXiv:hep-th/0609007]}.

M.~Sakaguchi and K.~Yoshida,
{\em A semiclassical string description of Wilson loop with local operators},
Nucl.\ Phys.\  B {\bf 798} (2008) 72, {\tt [arXiv:0709.4187 [hep-th]]}.

\bibitem{AT} L.~F.~Alday and A.~A.~Tseytlin,
{\em On strong-coupling correlation functions of circular Wilson loops and local operators},
J.\ Phys.\ A {\bf 44} (2011) 395401, 
{\tt [arXiv:1105.1537 [hep-th]]}.

R.~Hern\'andez, 
{\em Semiclassical correlation functions of Wilson loops and local vertex operators},
Nucl.\ Phys.\ B {\bf 862} (2012) 751, 
{\tt [arXiv:1202.4383 [hep-th]]}.

E.~I.~Buchbinder and A.~A.~Tseytlin,
{\em Correlation function of circular Wilson loop with two local operators and conformal invariance},
Phys.\ Rev.\ D {\bf 87} (2013) 026006, {\tt [arXiv:1208.5138 [hep-th]]}.

\bibitem{GiombiPestun} S.~Giombi and V.~Pestun,
{\em Correlators of Wilson loops and local operators from multi-matrix models and strings in AdS}, 
JHEP {\bf 1301} (2013) 101, {\tt [arXiv:1207.7083 [hep-th]]}.

\bibitem{Enari} T.~Enari and A.~Miwa,
{\em Semi-classical correlator for 1/4 BPS Wilson loop and chiral primary operator with large R-charge}, 
Phys.\ Rev.\ D {\bf 86} (2012) 106004, 
{\tt [arXiv:1208.0821 [hep-th]]}.

\bibitem{AHS} L.~F.~Alday, P.~Heslop and J.~Sikorowski,
{\em Perturbative correlation functions of null Wilson loops and local operators},
JHEP {\bf 1303} (2013) 074, {\tt [arXiv:1207.4316 [hep-th]]}.

\bibitem{AHS2} L.~F.~Alday, J.~M.~Henn and J.~Sikorowski,
{\em Higher loop mixed correlators in ${\cal N}=4$ SYM}, 
 JHEP {\bf 1303} (2013) 058, {\tt [arXiv:1301.0149 [hep-th]]}.

\bibitem{int} L.~F.~Alday, D.~Gaiotto and J.~Maldacena,
{\em Thermodynamic Bubble Ansatz},
JHEP {\bf 1109} (2011) 032, {\tt [arXiv:0911.4708 [hep-th]]}.

L.~F.~Alday, J.~Maldacena, A.~Sever and P.~Vieira,
{\em Y-system for Scattering Amplitudes},
J.\ Phys.\ A {\bf 43} (2010) 485401, {\tt [arXiv:1002.2459 [hep-th]]}.

\bibitem{hexagon} D.~Astefanesei, S.~Dobashi, K.~Ito and H.~Nastase,
{\em Comments on gluon 6-point scattering amplitudes in ${\cal N}=4$ SYM at strong coupling},
JHEP {\bf 0712} (2007) 077, {\tt [arXiv:0710.1684 [hep-th]]}.

\bibitem{ZaremboRT} K.~Zarembo,
{\em Holographic three-point functions of semiclassical states},
JHEP {\bf 1009} (2010) 030, {\tt [arXiv:1008.1059 [hep-th]]}.
  
M.~S.~Costa, R.~Monteiro, J.~E.~Santos and D.~Zoakos,
{\em On three-point correlation functions in the gauge/gravity duality},
JHEP {\bf 1011} (2010) 141, {\tt [arXiv:1008.1070 [hep-th]]}.
  
R.~Roiban and A.~A.~Tseytlin,  
{\em On semiclassical computation of 3-point functions of closed string vertex operators in $AdS_5 \times S^5$},
Phys.\ Rev.\ D {\bf 82} (2010) 106011, 
{\tt [arXiv:1008.4921 [hep-th]]}.

\bibitem{AMeven} L.~F.~Alday and J.~Maldacena,
{\em Minimal surfaces in AdS and the eight-gluon scattering amplitude at strong coupling},
{\tt [arXiv:0903.4707 [hep-th]]}.

L.~F.~Alday and J.~Maldacena,
{\em Null polygonal Wilson loops and minimal surfaces in Anti-de-Sitter space},
JHEP {\bf 0911} (2009) 082, {\tt [arXiv:0904.0663 [hep-th]]}.




\end{thebibliography}
\end{document}